\documentclass[prd,aps,floats,floatfix,nofootinbib,preprint]{revtex4-1}
\usepackage{amsmath}
\usepackage{amsfonts}
\usepackage{graphicx}
\usepackage{slashed}
\usepackage{dcolumn}
\usepackage{bm}
\usepackage{amssymb}
\usepackage{latexsym}
\usepackage{color}
\usepackage{longtable}

\definecolor{rossoCP3}{cmyk}{0,.88,.77,.40}
		\definecolor{graa}{rgb}{0.8,0.8,0.8}
		\definecolor{blaa}{rgb}{0.2,0.2,0.6}
\begin{document}

\title{\color{rossoCP3} S and T Parameters from a Light Nonstandard Higgs  \\ versus \\ Near Conformal Dynamics}

\author{Roshan Foadi$\color{rossoCP3} ^\spadesuit$}
\email[]{roshan.foadi@uclouvain.be}
\author{Francesco Sannino$\color{rossoCP3} ^\clubsuit$}
\email[]{sannino@cp3.dias.sdu.dk}
\affiliation{$\color{rossoCP3} ^\spadesuit$ Centre for Cosmology, Particle Physics and Phenomenology (CP3)
Chemin du Cyclotron 2, Universit\'e catholique de Louvain, Belgium \\
$\color{rossoCP3} ^\clubsuit${\color{rossoCP3} CP$^{~3}$- Origins} \& DIAS, University of Southern Denmark, Odense, Denmark }

\begin{abstract}
We determine the contribution to the $S$ and $T$ parameters coming from extensions of the standard model featuring a light nonstandard-like Higgs particle. We neatly separate, using the Landau gauge, the contribution from the purely nonstandard Higgs sector, from the one due to the interplay of this sector with the standard model. If the nonstandard Higgs sector derives from a new type of near conformal dynamics, the formalism allows to precisely link the intrinsic underlying contribution with the experimentally relevant parameters.
\\[.1cm]
{\footnotesize  \it Preprint: CP$^3$-Origins-2012-19 \& DIAS-2012-20}
\end{abstract}

\maketitle

\section{Introduction}

Uncovering the true mechanism behind electroweak symmetry breaking (EWSB) is the primary goal of the Large Hadron Collider (LHC) experiments. While waiting for the experimental verdict, a large variety of theoretical explanations have been proposed to date.  The Higgs sector of the standard model (SM) is one possibility. The latter, however, is well known to suffer from severe instability, as the mass of the Higgs boson is naturally driven to the ultraviolet cutoff scale. It is important to stress that a time honored mechanism for spontaneous symmetry breaking and stabilization of the symmetry breaking scale is well known to occur in Nature. Quantum chromo-dynamics (QCD) induces dynamically a chiral-symmetry-breaking quark condensate intimately related to the pion decaying constant, $f_\pi\simeq 93$ MeV. Because of this phenomenon, even in absence of the Higgs sector, the SM electroweak symmetry is already broken, and the gauge bosons acquire a small mass around $30$ MeV. It is therefore a fact that a {\it little} dynamical EWSB occurs in Nature already at a few tens of MeV's. However, the bulk of the gauge boson masses still needs to be explained. An interesting possibility is that a new strong interaction, technicolor (TC)~\cite{Weinberg:1979bn,Susskind:1978ms}, provides the bulk contribution to the EW gauge bosons masses: this occurs if the technipion decay constant is $F_T=F_\pi/\sqrt{N_D}$, where $F_\pi \simeq 246$ GeV, and $N_D$ is the number of electroweak technidoublets.

The SM fermions interact with the condensate, and acquire mass, through an additional interaction, extended technicolor (ETC) \cite{Dimopoulos:1979es,Eichten:1979ah}, which is spontaneously broken at a scale much larger than the EWSB scale. In addition to the SM fermion masses, the ETC interactions can potentially induce flavor-changing neutral currents (FCNC). Suppressing the latter requires raising the ETC scale, whereas generating the right masses (as well as mixings) in the SM fermion sector requires lowering the ETC scale, especially for the top quark. This tension can be alleviated if the TC dynamics together with the ETC interactions feels the potential presence, in the parameters of the theory, of a nearby infrared fixed point. \cite{Holdom:1981rm,Holdom:1984sk,Appelquist:1986an,Miransky:1988gk,Miransky:1996pd}. This is not an automatic feature of any strongly coupled theory, given that one can also have cases in which the presence of a nearby fixed point is not felt by the theory \cite{Sannino:2012wy}. The first kind of dynamics is known as walking\footnote{Here walking and near-conformal dynamics are meant as synonimous.} while the second as jumping  \cite{Sannino:2012wy}. Walking requires the underlying coupling to run slowly for a certain range of energies above the EWSB scale, while jumping will have a fast decreasing TC coupling constant even just above the EWSB scale. Walking dynamics is phenomenologically interesting in that it enhances the technifermion condensate, as the latter scales approximately with the anomalous dimension at the infrared fixed point, whereas the operators mediating FCNC are not enhanced. De facto, it separates the scale of flavor physics from the one responsible for EWSB. An example of controllable four-dimensional walking dynamics has only recently appeared in \cite{Antipin:2012kc} while lower dimensional field theories with similar dynamics have also been investigated in \cite{deForcrand:2012se,Nogradi:2012dj}.

Another interesting aspect of near-conformal TC is that, for smooth transitions between scale invariant and broken phase, one can expect a scalar isosinglet, call it $h$, to be light relative to the technihadron mass scale $M_\rho$. This is seen by comparing the trace anomaly in the underlying gauge theory -- in which it vanishes as the $\beta$ function approaches a fixed point -- with the trace anomaly in the effective theory, under the assumption that the latter is saturated by a single scalar field \cite{Sannino:1999qe,Hong:2004td,Dietrich:2005jn}. It is for this reason that such a scalar singlet is termed  {\em technidilaton}. It is natural to identify this state with the Higgs boson. In general this state can also have nonstandard couplings to the SM particles. The actual separation in mass between  the technidilaton and the remaining spectrum depends on the given model. We expect that a light dilatonic state, compared to the heaviest states of the theory, will appear in models featuring multiple couplings as shown in calculable examples \cite{Grinstein:2011dq,Antipin:2011aa}. Furthermore, by an explicit non-perturbative {\it exact} computation in \cite{Antipin:2011aa}, it was shown that the lightest states need not be the dilaton.

Whatever the underlying dynamics yielding a nonstandard light Higgs is, we assume its presence in the spectrum, and determine its consequences on precision electroweak observables. Because in the most general extensions of the SM the Higgs can have couplings to $WW$ and $ZZ$ differing from the SM one, the contributions to the Peskin-Takeuchi $S$ and $T$ parameters \cite{Peskin:1991sw} will not cancel against the SM contributions~\cite{Barbieri:2007bh}. Here we compute $S$ and $T$ by trying to disentangle the contributions which are due to the nonstandard Higgs sector (interpreted as coming from a new strongly coupled sector), from the contributions arising from the interplay between the strongly coupled sector and the electroweak sector. This can be achieved by employing the Landau gauge, in which the weak boson propagators are purely transverse (and thus correspond to elementary weak bosons) and the composite Goldstone bosons are massless (as in the strongly coupled sector, for zero electroweak gauge couplings). If the computation is done properly, the strong sector contribution to $T$ should vanish, as we only consider TC theories with custodial symmetry. On the other hand, the strong sector contribution to $S$ is not expected to vanish, but is arguably bounded by~\cite{Sannino:2010ca,Sannino:2010fh,DiChiara:2010xb}
\begin{equation}
S_{\rm naive} \leq S \lesssim 2S_{\rm naive} \ ,
\label{eq:bound}
\end{equation}
where $S_{\rm naive}$ is the one-loop contribution to the $S$ parameter in a weakly coupled regime. For a TC theory with $N_D$ left-handed electroweak technidoublets,
\begin{eqnarray}
S_{\rm naive} = \frac{N_D\ d(r)}{6\pi} \ ,
\end{eqnarray}
where $d(r)$ is the dimension of the technifermion representation. When comparing $S$ with experimental data we shall impose the bound of Eq.~(\ref{eq:bound}). The latter is in agreement with the expected reduction of the $S$ parameter in walking theories, due to the modification of the second Weinberg sum rule \cite{Appelquist:1998xf}. Assuming the bound implies that, if the Higgs contribution is large enough to push the estimate outside the $(S_{\rm naive},2S_{\rm naive})$ interval, the heavy resonance sector at the $M_\rho$ scale should adjust accordingly to restore the bound. Of course, our results at the effective Lagrangian level are general and do not depend on this bound. One can use directly the proper lattice determination of the electroweak parameters, coming from the intrinsic strong dynamics, and use our generalized results to link the latter to the experimentally relevant quantities. However, as we shall see, the precise match requires the knowledge of the three point function of a potentially light scalar state to two pions, as well as the light scalar mass. If these two quantities are hard to estimate on the lattice, then one can still estimate the uncertainty on the matching to experiments by varying these quantities with respect to the corresponding reference values of the SM Higgs.

In section \ref{nonlinear} we introduce a nonstandard Higgs-like state using nonlinear realizations, whereas in section \ref{precisions} we determine the $S$ and $T$ parameters. In particular, we disentangle the corrections coming from the strongly-coupled Higgs sector, from the ones deriving from the interplay of this sector with the SM.  We also elucidate the dependence of the $S$ and $T$ parameters upon the deviations of the nonstandard Higgs coupling to the SM gauge bosons. We compare with experiments in section \ref{exp} where we also offer our conclusions.

\section{The nonlinearly realized Higgs}
\label{nonlinear}
As motivated in the introduction, we consider theories in which the SM Higgs and the electroweak Goldstone bosons (i.e. the SM Higgs sector) is nonstandard. The nonstandard Higgs state is taken to be light with respect to the cutoff scale of the theory, indicated by $M_\rho$. This setup is general and can be reinterpreted as emerging, for example, from new strong dynamics in which the Higgs-like and the Goldstone states are composite. We assume the dynamics to be walking and that the composite Higgs state is light relative to the natural mass scale of the heavy resonances, i.e. $m_h\ll M_\rho$. At energies below $M_\rho$, the interactions of the composite Higgs with the SM particles are well described by effective Lagrangians in which all heavy resonances are integrated out. We assume that the new strong dynamics respects $SU(2)_c$ custodial isospin symmetry, and adopt a nonlinear realization for the composite states. The latter are thus classified according to linear multiplets of $SU(2)_c$: the electroweak Goldstone bosons form an $SU(2)_c$ triplet, whereas the Higgs is an $SU(2)_c$ singlet. To leading order in the momentum expansion, the interactions of the Higgs with the electroweak gauge and Goldstone bosons are therefore dominated by the Lagrangian term
\begin{equation}
{\cal L} = \frac{v^2}{4}\kappa\left[{h}/{v}\right]\ {\rm Tr} \left[ D_\mu U D^\mu U^\dagger \right] \ ,
\label{eq:L}
\end{equation}
where $U$ is, as usual, the exponential map of the Goldstone boson fields $\pi^a$,
\begin{equation}
U=\exp \left(2 i \pi^a T^a/v\right) \ , \quad a=1,2,3,
\end{equation}
$v\equiv 246$ GeV is the dynamically generated electroweak vev, and $2 T^a$ are the Pauli matrices. Adopting a noncanonical normalization, in which
the weak and hypercharge coupling constants, $g$ and $g^\prime$, respectively, are absorbed in the gauge fields, the covariant derivative of $U$ reads
\begin{equation}
D_\mu U = \partial_\mu U - i W^a_\mu T^a U + i U B_\mu T^3 \ .
\end{equation}
Since the Higgs field $h$ is an $SU(2)_c$ singlet, the function $\kappa$ in Eq.~(\ref{eq:L}) is arbitrary. The interaction vertices are obtained by expanding $\kappa$ around $h/v=0$, with higher order terms in the expansion being suppressed by inverse powers of the cutoff scale. A proper normalization of the Goldstone boson kinetic terms requires
\begin{equation}
\kappa[0] = 1 \ .
\end{equation}
We also define
\begin{equation}
\kappa_1 \equiv \kappa^\prime[0]\ , \quad
\kappa_2 \equiv \kappa^{\prime\prime}[0] \ ,
\end{equation}
where the derivative is respect to the argument of $\kappa$, i.e. $h/v$. The nonlinearly realized SM is obtained by setting
\begin{equation}
\kappa^{\rm SM}[h/v] = \left(1+\frac{h}{v}\right)^2 \ ,
\end{equation}
whence
\begin{equation}
\kappa_1^{\rm SM} = \kappa_2^{\rm SM} = 2 \ .
\end{equation}
The interaction vertices of one and two Higgs bosons with the electroweak gauge and Goldstone bosons are therefore obtained by rescaling the SM vertices by $\kappa_1/\kappa_1^{\rm SM}$, and $\kappa_2/\kappa_2^{\rm SM}$, respectively.

%
%
%
%
\section{Nonstandard Light Higgs Contribution to $S$ and $T$}
\label{precisions}
In this section we compute the contribution of the light composite Higgs to the Peskin-Takeuchi $S$ and $T$ parameters~\cite{Peskin:1991sw}. It is convenient to do so in Landau gauge, as this marks a clear distinctions between the elementary and transversely polarized electroweak bosons, and the massless composite Goldstone bosons.
\subsection{$S$ Parameter}
\begin{figure}
\includegraphics[width=5.0in]{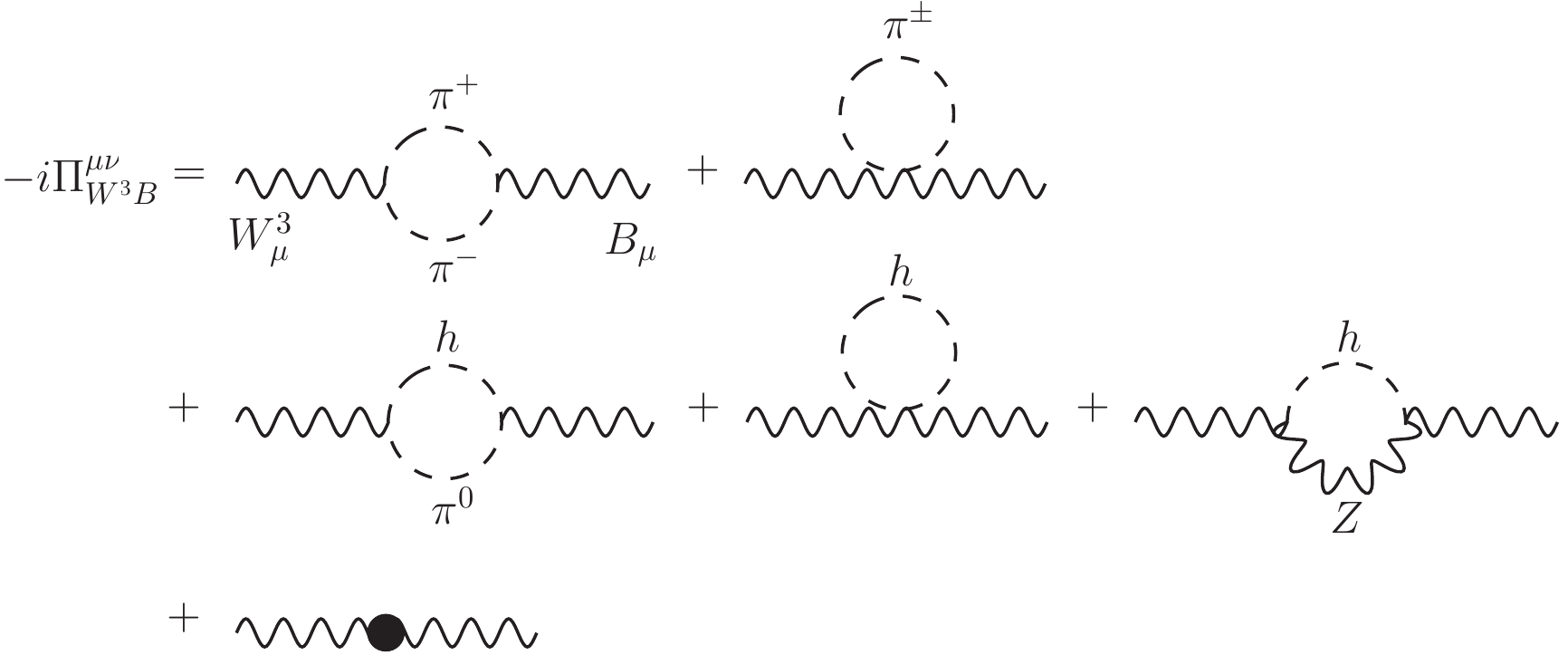}
\caption{One loop diagrams for the Goldstone boson and Higgs boson contributions to $\Pi_{W^3 B}^{\mu\nu}(q)$. The last diagram is the counterterm, which depends on the dynamics at and above the cutoff scale $M_\rho$. In Landau gauge, all diagrams except the $hZ$ exchange arise entirely from the strongly interacting sector.}
\label{fig:W3B}
\end{figure}
The one-loop contribution of the composite Higgs and Goldstone bosons to the $\Pi_{W^3 B}^{\mu\nu}(q)$ correlator is given by the first five diagrams of Fig.~\ref{fig:W3B}. The last diagram is the counterterm, which depends on the dynamics at and above the cutoff scale $M_\rho$. Define $\Pi_{W^3 B}(q^2)$ as the coefficient of $g^{\mu\nu}$ in $\Pi_{W^3 B}^{\mu\nu}(q)$. This turns out to be quadratically divergent, unless $\kappa=\kappa^{\rm SM}$. However the quadratic divergence drops out in $\Pi_{W^3 B}^\prime(q^2)\equiv d\Pi_{W^3 B}(q^2)/dq^2$, which is the quantity of interest for us. Thus we can employ dimensional regularization and set\footnote{The log argument becomes dimensionless when combined with masses and momenta in the loop.}
\begin{equation}
\frac{1}{\epsilon}-\gamma+\log 4\pi = \log M_\rho^2 \ .
\label{eq:inf}
\end{equation}
Computing the diagrams of Fig.~\ref{fig:W3B} gives
\begin{eqnarray}
&& \Pi_{W^3 B}^\prime(q^2) = \frac{1}{16\pi^2}\left[\frac{1}{12}\log\frac{M_\rho^2}{m_\pi^2-q^2}+\frac{5}{36}\right]
+\frac{1}{16\pi^2}\frac{\kappa_1^2}{4}\int_0^1 dx \frac{x(1-x)}{2}\log\frac{M_\rho^2}{x\ m_h^2-x(1-x)q^2} \nonumber \\
&+&\frac{1}{16\pi^2}\frac{\kappa_2}{8}\int_0^1 dx \left[-x(2-3x)\log\frac{m_h^2}{x\ m_h^2-x(1-x)q^2}
+x(1-x)\left(1+\frac{x\ q^2}{x\ m_h^2-x(1-x)q^2}\right)\right] \nonumber \\
&+&\frac{1}{16\pi^2}\frac{\kappa_1^2}{4}\int_0^1 dx \int_0^{1-x} dy \Big[
\frac{x\ m_Z^2}{x\ m_h^2+(1-x)m_Z^2-x(1-x)q^2}
-\frac{x(1-x) m_Z^2/2}{x\ m_h^2+y\ m_Z^2-x(1-x)q^2}\Big]
+ c(M_\rho) \ , \nonumber \\
\label{eq:W3B}
\end{eqnarray}

where the tree-level counterterm is obtained by including the ${\cal O}(p^4)$ and $SU(2)_c$-symmetric operator
\begin{eqnarray}
-\frac{c(M_\rho)}{g g^\prime} \ {\rm Tr}\ W^a_{\mu\nu} T^a U B^{\mu\nu} T^3 U^\dagger \ ,
\end{eqnarray}
to the Lagrangian. This counterterm was introduced and discussed in the classic paper by Gasser and Leutwyler in \cite{Gasser:1983yg} and used for traditional approaches to TC, for example, in \cite{Dobado:1990zh,Pich:2012jv,Bijnens:2009qm}.

The first two terms between square brackets correspond to the Goldstone boson contribution, that is, the first two diagrams in Fig.~\ref{fig:W3B}. Notice that a small mass for the Goldstone bosons has been added, in order to regulate the corresponding infrared divergence. The first two integrals correspond to the third and fourth diagrams, respectively, in Fig.~\ref{fig:W3B}. Such terms arise entirely from the strong sector. Finally, the double integral is due to the interaction between the composite Higgs and the transverse elementary electroweak bosons, and correspond to the fifth diagram in Fig.~\ref{fig:W3B}. Such a contribution is finite and unambiguous, and should be added to the contribution arising from the strong sector.

The ``total'' $S$ parameter (without subtraction of the corresponding SM contribution) is
\begin{eqnarray}
S_{\rm tot} = 16\pi \Pi_{W^3 B}^\prime(0) &=& \frac{1}{12\pi}\log\frac{M_\rho^2}{m_\pi^2}+\frac{5}{36\pi}
-\frac{\kappa_1^2}{4}\left[\frac{1}{12\pi}\log\frac{M_\rho^2}{m_h^2}
+\frac{5}{72\pi}+\frac{f(m_Z/m_h)}{6\pi}\right] \nonumber \\
&+& 16\pi\ c(M_\rho) \ ,
\label{eq:Stot}
\end{eqnarray}
where
\begin{equation}
f(x)\equiv\frac{2x^2+x^4-3x^6+(9x^4+x^6)\log x^2}{(1-x^2)^3} \ .
\end{equation}
The first two terms between square brackets correspond to the third diagram in Fig.~\ref{fig:W3B}, whereas the last term corresponds to the fifth diagram. The seagull diagram, proportional to $\kappa_2$, gives no contribution at zero momentum. Finally, the counterterm is made of two distinct parts: a numerical part proportional to $\kappa_1$, and a part which vanishes in the limit $M_\rho\to\infty$. The latter arises from the resonances at and above the $M_\rho$ scale, and goes to zero as $M_\rho\to\infty$. The former is scheme dependent, and is expected to be as small as the non-log term in the Higgs loop.

The SM contribution, computed at a reference mass $m_{h,{\rm ref}}$, is
\begin{eqnarray}
S_{\rm SM} &=&  \frac{1}{12\pi}\log\frac{M_\rho^2}{m_\pi^2}
-\frac{1}{12\pi}\log\frac{M_\rho^2}{m_{h,{\rm ref}}^2}+\frac{5}{72\pi}- \frac{f(m_Z/m_{h,{\rm ref}})}{6\pi} \ ,
\label{eq:SSM}
\end{eqnarray}
where no counterterm is needed, since the ultraviolet divergence arising from the Goldstone boson loop cancels the log divergent term arising from the Higgs-Goldstone exchange. The $S$ parameter is given by $S_{\rm tot}-S_{\rm SM}$, whence
\begin{eqnarray}
S &=& \left(1-\frac{\kappa_1^2}{4}\right)\left[\frac{f(m_Z/m_h)}{6\pi}+\frac{1}{12\pi}\log\frac{M_\rho^2}{m_h^2}
+\frac{5}{72\pi}\right] + 16\pi\ c(M_\rho) \nonumber \\
&+& \frac{1}{12\pi}\log\frac{m_h^2}{m_{h,{\rm ref}}^2} + \frac{f(m_Z/m_{h,{\rm ref}})-f(m_Z/m_h)}{6\pi}\ .
\label{eq:S0}
\end{eqnarray}
This quantity should be compared with the experimental bounds obtained for the Higgs mass equal to its reference value.  The first term is a direct measure of the deviation, from the SM value, of the Higgs coupling to the weak bosons. This was not considered in \cite{Peskin:1991sw}.

In Eq.~(\ref{eq:S0}) the log tells us that for $\kappa_1<2$ ($\kappa_1>2$) there is a positive (negative) contribution to the $S$ parameter proportional to the large log which separates the $M_\rho$ scale from the $m_h$ scale. However we should keep in mind that the contribution from the strong dynamics, $S_{\rm strong}$, is expected to be bounded as in Eq.~(\ref{eq:bound}). In Eq.~(\ref{eq:Stot}) $S_{\rm strong}$  is given by all terms except the last one between square brackets,
\begin{eqnarray}
S_{\rm strong} = \left(1-\frac{\kappa_1^2}{4}\right)\left[\frac{1}{12\pi}\log\frac{M_\rho^2}{m_h^2}+\frac{5}{72\pi}\right]
+\frac{5}{72\pi}+\frac{1}{12\pi}\log\frac{m_h^2}{m_\pi^2}  + 16\pi\ c(M_\rho) \ .
\end{eqnarray}

For $\kappa_1<2$ the light composite Higgs gives a positive contribution, reinforcing the existence of a lower bound for $S_{\rm strong}$~\cite{Sannino:2010ca,Sannino:2010fh,DiChiara:2010xb}. Combining the last two equations yields
\begin{eqnarray}
S &=& S_{\rm strong} + \left(1-\frac{\kappa_1^2}{4}\right)\frac{f(m_Z/m_h)}{6\pi} - \frac{5}{72\pi}- \frac{1}{12\pi}\log\frac{m_h^2}{m_\pi^2} \nonumber \\
&+& \frac{1}{12\pi}\log\frac{m_h^2}{m_{h,{\rm ref}}^2} + \frac{f(m_Z/m_{h,{\rm ref}})-f(m_Z/m_h)}{6\pi}\ .
\label{eq:S}
\end{eqnarray}
Notice that in the strong-sector contribution to the $S$ parameter the infrared divergence arising from the Goldstone boson loop is subtracted. This is equivalent to taking $m_\pi$ of the order of $m_h$ or $v$, in such a way that no large logs appear in the final result.
\subsection{$T$ Parameter}
\begin{figure}
\includegraphics[width=6.0in]{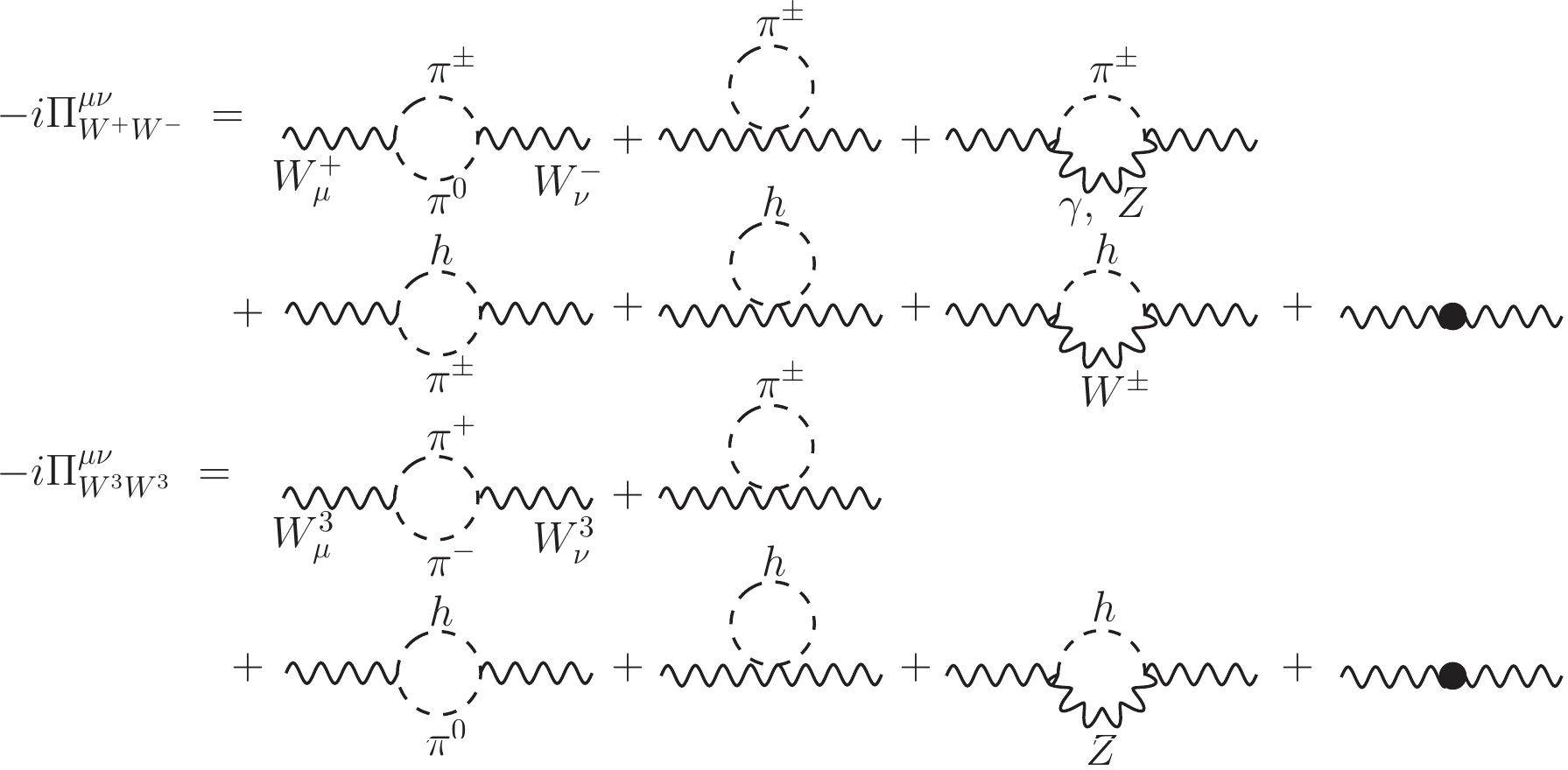}
\caption{One loop diagrams for the Goldstone boson and Higgs boson contributions to $\Pi_{W^+ W^-}^{\mu\nu}(q)$ and $\Pi_{W^3 W^3}^{\mu\nu}(q)$. The last diagram is the counterterm, which depends on the dynamics at and above the cutoff scale $M_\rho$.}
\label{fig:WW}
\end{figure}
The contribution of the composite Higgs and Goldstone exchanges to $\Pi_{W^+ W^-}^{\mu\nu}(q)$ and $\Pi_{W^3 W^3}^{\mu\nu}(q)$ is given by the diagram of Fig.~\ref{fig:WW}. Define the coefficient of $g^{\mu\nu}$, in $\Pi_{W^+ W^-}^{\mu\nu}(q)$ and $\Pi_{W^3 W^3}^{\mu\nu}(q)$, as $\Pi_{W^+ W^-}(q^2)$ and $\Pi_{W^3 W^3}(q^2)$, respectively. The $T$ parameter is proportional to the difference $\Pi_{W^3 W^3}(0)-\Pi_{W^+ W^-}(0)$, which is free of quadratic divergences: thus we can employ dimensional regularization and interpret the infinity as the log of the cutoff $M_\rho$, as in Eq.~(\ref{eq:inf}). In Landau gauge the only diagrams which do not cancel in $\Pi_{W^3 W^3}(0)-\Pi_{W^+ W^-}(0)$ are those with the exchange of a weak boson and either a Higgs or a Goldstone boson. These do not arise from the new strong sector, but rather from the interaction of the latter with the electroweak sector. This is consistent with the fact that the underlying dynamics preserves $SU(2)_c$, and explicitly shows the disentangling of the electroweak sector and the composite sector in Landau gauge: had we employed the 't Hooft-Feynman gauge, for instance, the Goldstone-Goldstone exchanges would have given a nonzero contribution to $\Pi_{W^3 W^3}(0)-\Pi_{W^+ W^-}(0)$.  An explicit computation gives
\begin{eqnarray}
&& \Pi_{W^3 W^3}(q^2)-\Pi_{W^+ W^-}(q^2) = \Pi^\pi_{W^3 W^3}(q^2)-\Pi^\pi_{W^+ W^-}(q^2)
+\frac{1}{16\pi^2}\frac{\kappa_1^2}{4}\frac{3}{4}(m_Z^2-m_W^2)\log\frac{M_\rho^2}{m_h^2} \nonumber \\
&+&\frac{1}{16\pi^2}\frac{\kappa_1^2}{4}\int_0^1 dx \int_0^{1-x} dy \Big[
\frac{m_Z^2}{1-x}\log\frac{m_h^2}{x m_h^2+(1-x)m_Z^2-x(1-x)q^2} - \frac{m_Z^2}{2}\log\frac{m_h^2}{x m_h^2+y\ m_Z^2-x(1-x)q^2} \nonumber \\
&+&\frac{m_W^2}{1-x}\log\frac{m_h^2}{x m_h^2+(1-x)m_W^2-x(1-x)q^2} + \frac{m_W^2}{2}\log\frac{m_h^2}{x m_h^2+y\ m_W^2-x(1-x)q^2}\Big]
+ \frac{v^2}{4}c^\prime(M_\rho) \ ,
\end{eqnarray}
where $\Pi^\pi_{W^3 W^3}(q^2)-\Pi^\pi_{W^+ W^-}(q^2)$ is the contribution from the Goldstone-gauge exchanges, and the tree-level counterterm diagram is obtained by including the ${\cal O}(p^2)$ and $SU(2)_c$-violating operator
\begin{eqnarray}
-\frac{v^2}{2}c^\prime(M_\rho)\left({\rm Tr}\ T^3 U^\dagger D_\mu U \right)^2
\end{eqnarray}
to the Lagrangian. The total contribution to $T$ is
\begin{eqnarray}
T_{\rm tot} &=& \frac{4}{\alpha\ v^2} \left[\Pi_{W^3 W^3}(0)-\Pi_{W^+ W^-}(0)\right] = T^\pi
+\frac{\kappa_1^2}{4} T_{\rm SM}^h(m_h) +\frac{c^\prime(M_\rho)}{\alpha}  \ ,
\end{eqnarray}
where $T^\pi$ is the contribution from the Goldstone-gauge exchanges, and $T_{\rm SM}^h(m_h)$ is the (divergent) SM Higgs contribution:
\begin{eqnarray}
T_{\rm SM}^h(m_h) &=& \frac{3}{16\pi c^2}\Bigg[\log\frac{M_\rho^2}{m_h^2}
+\frac{5}{6}-\frac{1}{s^2}\frac{m_Z^2\log(m_h^2/m_Z^2)}{m_h^2-m_Z^2}
+\frac{c^2}{s^2}\frac{m_W^2\log(m_h^2/m_W^2)}{m_h^2-m_W^2}\Bigg] \ .
\label{eq:TSMh}
\end{eqnarray}
Here  $c$ and $s$ are the cosine and sine of the weak mixing angle and should not be confused with the counterterms always indicated as functions of $M_\rho$.  In the SM the contribution to $T$ at the Higgs reference mass is
\begin{eqnarray}
T_{\rm SM} &=& T^\pi +  T_{\rm SM}^h(m_{h,{\rm ref}}) \ ,
\end{eqnarray}
where no counterterm is needed, as the infinite terms cancel between $T^\pi$ and $T_{\rm SM}^h(m_{h,{\rm ref}})$. The $T$ parameter is given by the difference $T_{\rm tot}-T_{\rm SM}$. Since $T^\pi$ is identical in the theory with a light composite Higgs and in the SM, this gives
\begin{eqnarray}
T &=& -\left(1-\frac{\kappa_1^2}{4}\right)T_{\rm SM}^h(m_h) + T_{\rm SM}^h(m_h) - T_{\rm SM}^h(m_{h,{\rm ref}}) + \frac{c^\prime(M_\rho)}{\alpha} \ .
\label{eq:T0}
\end{eqnarray}
As in the contribution to the $S$ parameter, the counterterm consists of a part of the same size of the finite terms, in Eq.~(\ref{eq:TSMh}), and a part which arises from the resonances at and above the $M_\rho$ scale. The latter is approximately zero, in a theory with custodial symmetry. The former, together with the finite terms in Eq.~(\ref{eq:TSMh}), can be absorbed in the definition of $M_\rho$. This leads to the final expression for $T$:
\begin{eqnarray}
T &=& -\left(1-\frac{\kappa_1^2}{4}\right)\log\frac{M_\rho^2}{m_h^2} + T_{\rm SM}^h(m_h) - T_{\rm SM}^h(m_{h,{\rm ref}}) \ .
\label{eq:T}
\end{eqnarray}
For $\kappa_1<2$ ($\kappa_1>2$) we obtain a negative (positive) contribution proportional to $\log M_\rho^2/m_h^2$: as above mentioned, such a contribution does not arise from the new strong dynamics, and should be added to the latter.
\section{Comparison with Experiment and Conclusions}
\label{exp}
\begin{figure}
\includegraphics[width=6in]{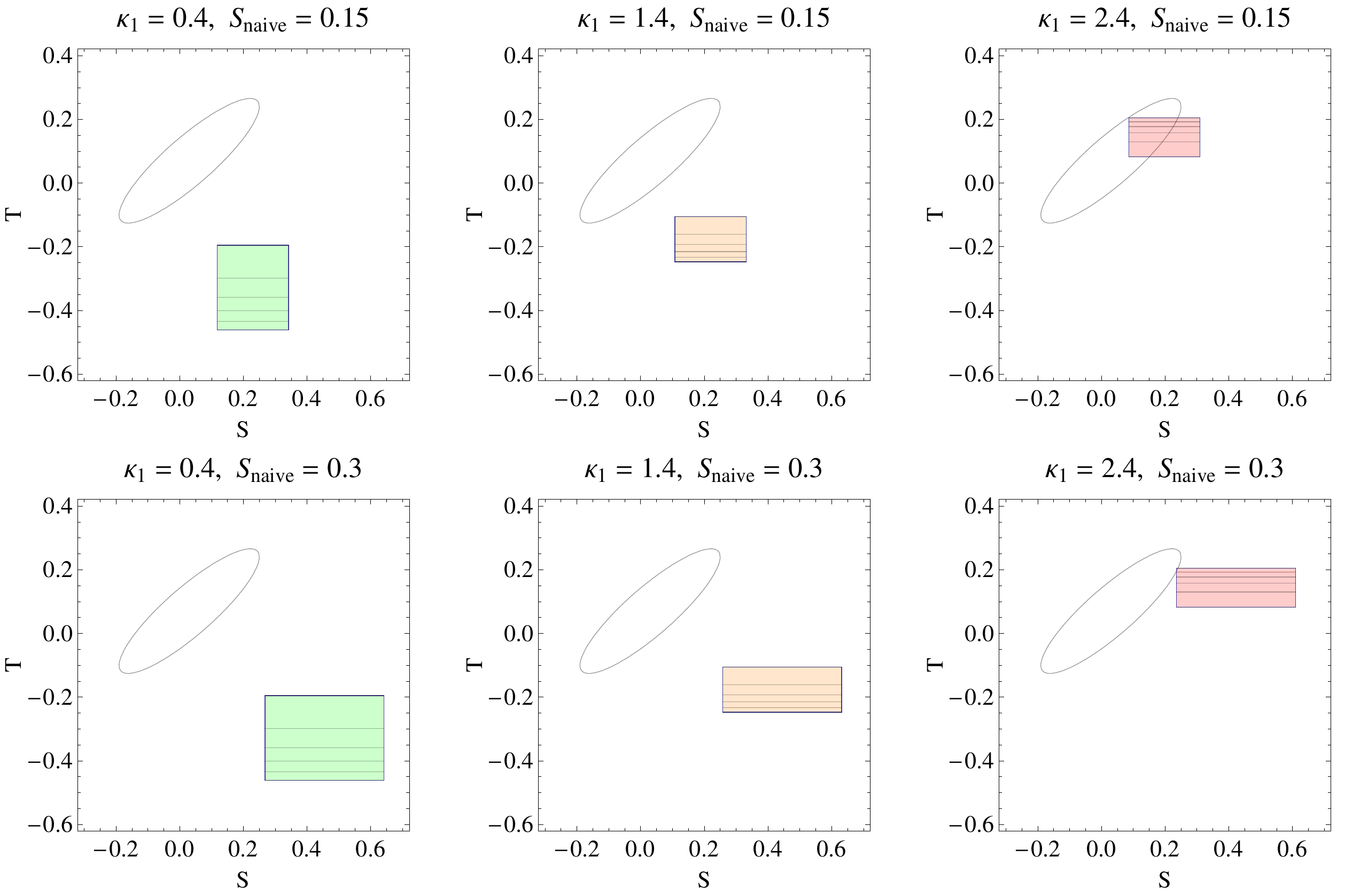}
\caption{}
\label{fig:ST}
\end{figure}
Here we would like to use the determination of $S$ and $T$ from Eqs.~(\ref{eq:S}) and (\ref{eq:T}), respectively, and compare these with the experimental constraints. The determination of $S$ has a large source of uncertainty given by the bound of Eq.~(\ref{eq:bound}), and a smaller source given by the value of $m_\pi$ employed to remove the infrared divergence: here we set $m_h/2 < m_\pi < 2 m_h$. The uncertainty in the determination of $T$ is due to the counterterm, and to the precise value of $T$ at the scale $M_\rho$. The latter can be assumed to vanish, due to the $SU(2)_c$ symmetry. The former is absorbed in the definition of the logarithmic cutoff $\log(M_\rho^2/m_h^2)$. Therefore, for each value of $\kappa_1$ and $S_{\rm naive}$ the TC prediction is a vertical band in the $(S,T)$ plane, as shown in Fig.~\ref{fig:ST}. Within each band, the horizontal lines correspond to $M_\rho=0.5, 1.0, 1.5, 2.0, 2.5, 3.0$ TeV. Within each plot we also show the 95\% C.L. LEP bounds for $m_{h,{\rm ref}}=117$ GeV. Notice that for $\kappa_1<2$ the TC contribution to $S$ ($T$) is too large (small) by a few standard deviations. Additional contributions arise from the ETC sector, and from new sectors of the theory which do not participate in the strong dynamics. For example, in Ref.~\cite{Foadi:2007ue} it was shown that the extra lepton doublet of minimal walking technicolor, introduced to cure the topological $SU(2)_w$ Witten anomaly, can be adjusted to give negative (positive) contributions to $S$ ($T$), restoring agreement with the experimental bounds. The agreement is even improved, if the negative contribution to $T$ from the Higgs is taken into account. For $\kappa_1>2$ the TC contribution agrees better with the experimental bounds. Notice, however, that a $g_{hWW}$ coupling significantly larger than $g_{hWW}^{\rm SM}$ is disfavored by unitarity arguments, if the longitudinal $WW$ scattering amplitude is {\em solely} unitarized, at a few TeVs, by $h$ and a technirho. In fact the Higgs and the technirho both give positive contributions proportional to $E^2$, where $E$ is the center-of-mass energy, and this induces a unitarity loss already at a very few TeVs, if $g_{hWW}>g_{hWW}^{\rm SM}$\cite{Foadi:2008xj}. It is of course possible that an additional resonance, heavier than the technirho, restores unitarity by providing the necessary flip in sign to the term growing like $E^2$: this is indeed what occurs to the pion-pion scattering amplitude in QCD, in which the $f_0(980)$ scalar resonance overturns the negative contributions from the broad sigma and the rho meson~\cite{Harada:1995dc}.

\vskip .5cm
We  determined the contribution to precision observables coming from extensions of the SM  featuring a light nonstandard-like Higgs particle. Using the Landau gauge we disentangled the contribution of the nonstandard Higgs sector from the one due to the interplay of this sector with the SM. To compare with the experiment we then assumed that the nonstandard Higgs sector derives from a new type of near conformal dynamics. We were able to get a better understanding of the size of the corrections coming from the presence of light Higgs-like state. We also showed that the formalism allows to link, in a precise manner, the contribution to the $S$ and $T$ parameters coming from the underlying dynamics with the experimentally relevant parameters. Our results, as explained in the introduction, are of direct interest to current lattice simulations of strongly coupled dynamics.

\end{document}